\documentclass{PoS}
\usepackage[utf8]{inputenc}

\title{ChPT meets lattice: finite volume and partial quenching for masses,
 decay constants and VEVs at NNLO}

\ShortTitle{Finite volume for masses and decay constants}

\author{\speaker{Thomas Rössler}\\
        {Department of Astronomy and Theoretical Physics, Lund University,\\
S\"olvegatan 14A, SE 223-62 Lund, Sweden}\\
        E-mail: \email{tr@thep.lu.se}}

\author{Johan Bijnens\\
        {Department of Astronomy and Theoretical Physics, Lund University,\\
S\"olvegatan 14A, SE 223-62 Lund, Sweden}\\
        E-mail: \email{bijnens@thep.lu.se}}

\abstract{We discuss finite volume effects and partial quenching for
Chiral Perturbation Theory in the mesonic sector. The effects are computed in
terms of analytical expressions for masses, decay constants and vacuum
expectation values (VEVs) to two-loop order. Numerical examples are presented
for a number of interesting and relevant cases. All numerical programs are
publically available, the prospects of a combination of these studies with
lattice gauge theory computations are discussed briefly.}

\FullConference{The 8th International Workshop on Chiral Dynamics, CD2015 ***\\
		29 June 2015 - 03 July 2015\\
		Pisa, Italy}

\renewcommand{\logo}
{\raisebox{-5mm}
{\parbox{\textwidth}{\begin{flushright}
LU TP 15-53\\
November 2015\\
\end{flushright}
\includegraphics{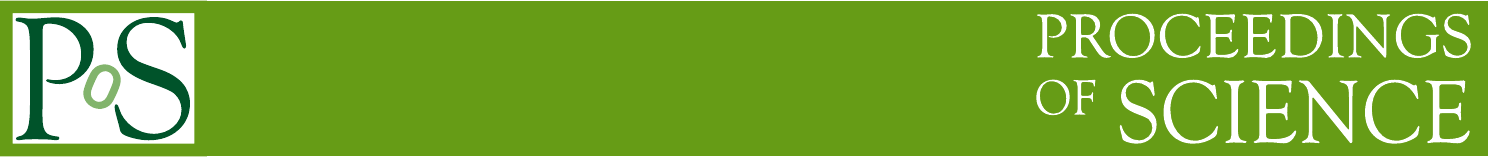}}
}}
\FullConference{{\rm To be published in the proceedings of:\\}
The 8th  International Workshop on Chiral Dynamics \\
  29 June 2015 - 03 July 2015\\
  Pisa, Italy}
\begin{document}

\section{Introduction: Chiral Perturbation Theory and unphysical artefacts}

Chiral Perturbation Theory (ChPT)
\cite{Weinberg:1978kz,Gasser:1983yg,Gasser:1984gg}
is the effective field theory for QCD
when looking at physics in the low-energy limit. Unphysical effects and
artefacts as needed in lattice gauge theory can be systematically included
in this framework. Calculations in Lattice Gauge Theory can highly benefit
when ChPT-guided extrapolations are employed to keep these effects - such as
unphysical valence and/or sea quark masses, the finite lattice spacing, the
finite lattice size and lattice artefacts, due to the choice of action, 
under control.

The talk discusses progress for two of these effects inside the ChPT
framework: the finite size and the unphysical masses
\cite{Bijnens:2014dea,Bijnens:2015dra}. To simplify the use
for the lattice community, we provide fully flexible and unrestricted
access to our numerical programs via the CHIRON \cite{Bijnens:2014gsa}
program collection. This can be downloaded from \cite{chiron}.
The analytical expressions can be downloaded from \cite{chptweb}.

It is inevitably the nature of any numerical lattice calculation that it
is performed in a finite volume. In order to perform a proper matching between
a lattice QCD calculation and ChPT, good control over the additional
quantum corrections to physical quantities which emerge due to the finiteness
of the volume is required. Then, infinite volume ChPT can serve as a
reliable validity check for the lattice calculation, or low-energy constants
(LECs) can be properly extracted from the lattice result.
Conceptually, in our work, a finite volume (FV) is introduced that restricts the
size of the Euclidean 3-dimensional space. The ``time'' dimension is assumed
to be infinitely extended. This treatment clearly breaks Lorentz invariance.
Dealing with the effects of all of this in ChPT at two-loop order is quite
involved. We discuss first shortly finite volume loop integrals,
Sect.~\ref{sec:FV}, then our results for standard ChPT in finite volume
\cite{Bijnens:2014dea} in Sect.~\ref{sec:FVChPT}. The largest part is devoted
to the partially quenched case, Sect.~\ref{sec:FVPQChPT}. We present
a few numerical results as well as the checks performed. Finally,
we point out the recent extension to QCD-like theories.

Introductions to ChPT can be found in the talks by Ecker \cite{Ecker:2015uoa}
and Bernard \cite{Bernard:2015wda} at this conference.
A more extensive introduction aimed at lattice
theorists is \cite{Golterman:2009kw}.
 
\section{Finite volume integrals}
\label{sec:FV}

Two different methods for the numerical evaluation of FV integrals have been
suggested at one-loop order
\cite{Gasser:1986vb,Gasser:1987ah,Gasser:1987zq,Becirevic:2003wk}. 
The integrals over momentum components in finite size direction are really
discrete sums. Using the Poisson summation theorem
the sum can be turned into a sum over integrals again.
The advantage of doing this is that the infinite volume expression can easily
be removed from this. The separation of the infinite volume (IV) and FV part
of an integral leaves a sum of integrals as the structure to be calculated
numerically. Depending on the preferred method, either the summation or the
integration can be eliminated. The actual evaluation routines then have to
solve only a summation over modified Bessel functions
\cite{Gasser:1986vb,Gasser:1987ah,Gasser:1987zq} or a numerical integration
over Jacobi theta functions \cite{Becirevic:2003wk}. This is strictly true for
the one-propagator cases, for two or more propagators additional numerical
integrations over Feynman parameters might be necessary.
Note that since the finite volume breaks Lorentz invariance extra structures
can appear in the integrals.

The extension of both methods to the general two-loop
sunset integral can be found in \cite{Bijnens:2013doa}. Expressions
for all the integrals needed in this work can be found there.
They are also available in the program package
CHIRON \cite{Bijnens:2014gsa,chiron}. The precise definitions
for the Minkowski versions we use in our work discussed
here can be found in \cite{Bijnens:2014dea,Bijnens:2015dra}.

\section{Finite volume for standard ChPT}
\label{sec:FVChPT}

\begin{figure}[tb]
\begin{center}
\includegraphics[width=0.5\textwidth]{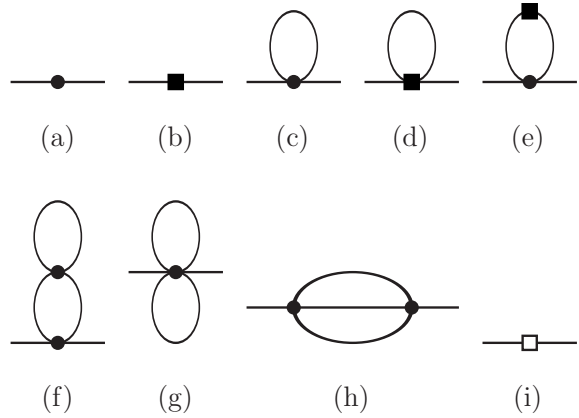}
\end{center}
\caption{The Feynman diagrams needed for the mass calculation. 
A dot indicates a vertex of order $p^2$, a filled box a vertex of order $p^4$
and an open box a vertex of order $p^6$.}
\label{figdiagrams}
\end{figure}

The calculation at finite volume is very much like the calculations
in infinite volume but one has to keep in mind the extra terms and extra
structures in the loop-integrals.
The diagrams that need to be evaluated are depicted in Fig.~\ref{figdiagrams}.

For numerical inputs we use
\begin{eqnarray}
\label{IVnuminputs}
F_\pi&=&92.2~\mathrm{MeV},~m_\pi=134.9764~\mathrm{MeV},
\mu = 770~\mathrm{MeV},~m_K=494.53~\mathrm{MeV},
\nonumber\\
m_\eta&=&547.30~\mathrm{MeV},~
\overline l_1 = -0.4,~\overline l_2=4.3,~
\overline l_3=3.0\,~\overline l=4.3\,.
\end{eqnarray}
For the three flavour LECs ($L^r_i$) we use the values
of the recent fit to continuum data \cite{Bijnens:2014lea}.

The analytical result for the finite volume correction to the pion mass
for the two flavour case agrees with the one-loop result of \cite{Gasser:1987zq}
and as far as we were able to check with the two-loop result of
\cite{Colangelo:2006mp}. The infinite volume result agrees with
\cite{Burgi:1996qi,Bijnens:1995yn}. 
For the three flavour case we agree for all masses with the infinite
volume result \cite{Amoros:1999dp} and the one-loop finite volume results
\cite{Becirevic:2003wk}.

\begin{figure}[tb]
\begin{minipage}{0.49\textwidth}
\includegraphics[width=0.99\textwidth]{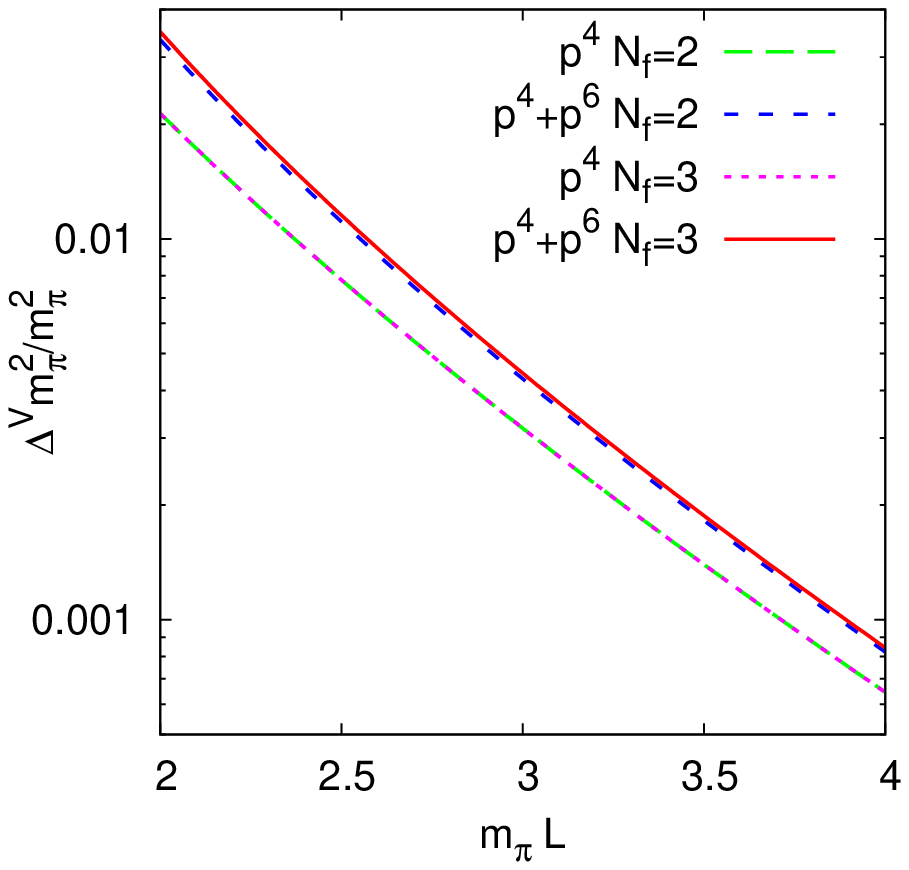}
\vskip-5mm
\centerline{(a)}
\end{minipage}
\begin{minipage}{0.49\textwidth}
\includegraphics[width=0.99\textwidth]{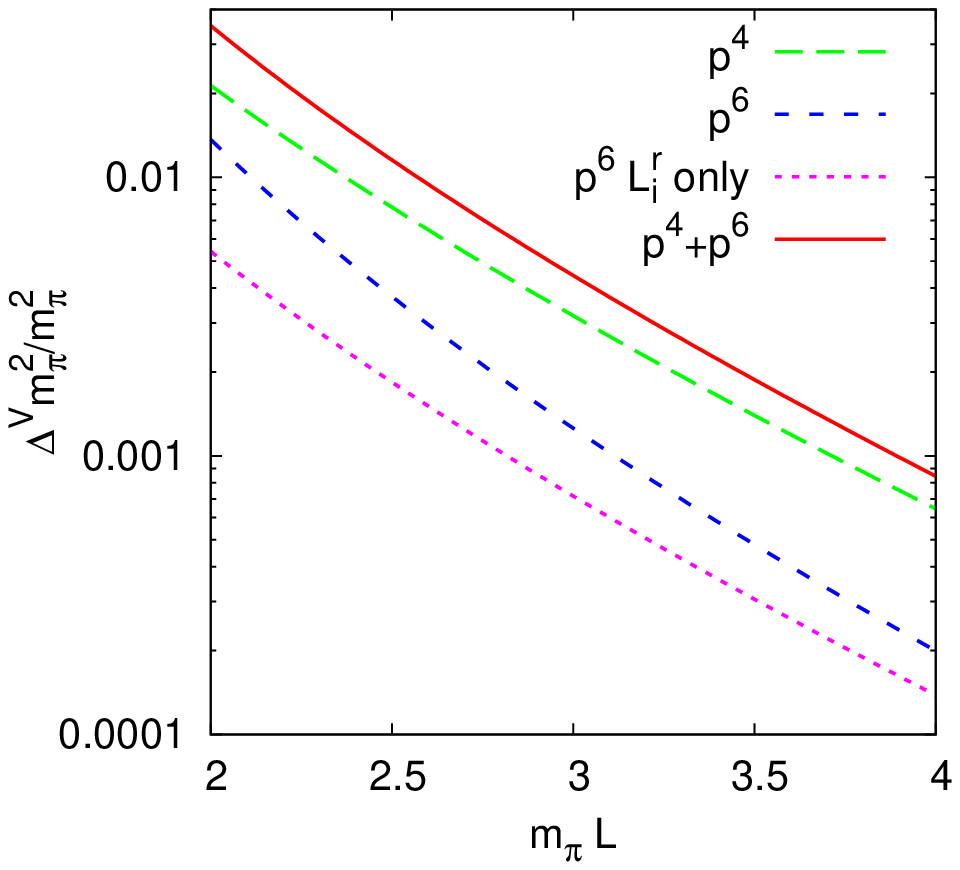}
\vskip-5mm
\centerline{(b)}
\end{minipage}
\caption{The relative finite volume corrections to the pion mass squared
with varying $L$. The other inputs are given in the main text.
(a) Comparison of the two- and three-flavour results. 
(b) The three-flavour case also showing
the $L_i^r$ dependent part.}
\label{figmpi}
\end{figure}

Numerical results for the pion mass are shown in Fig.~\ref{figmpi}.
We have varied $L$ but kept all other inputs constant as given
in (\ref{IVnuminputs}). The two and three flavour results are numerically in
good agreement showing that as expected the kaon and eta effects are small.

\begin{figure}[tb]
\begin{minipage}{0.49\textwidth}
\includegraphics[width=0.99\textwidth]{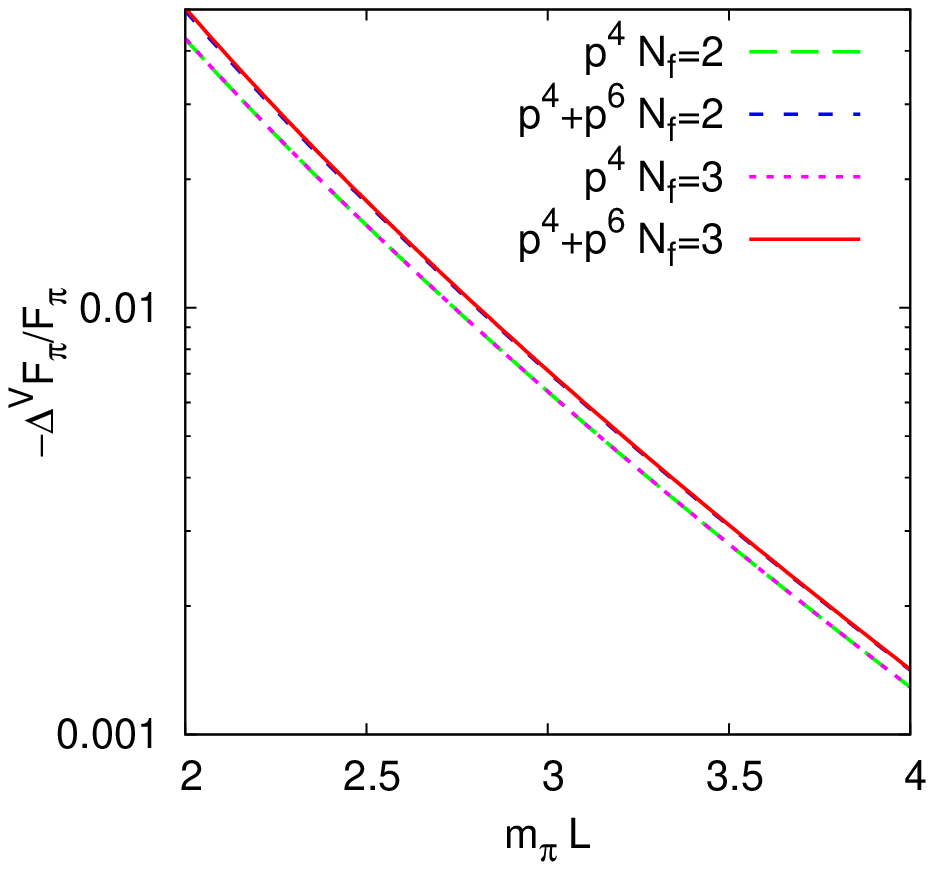}
\vskip-5mm
\centerline{(a)}
\end{minipage}
\begin{minipage}{0.49\textwidth}
\includegraphics[width=0.99\textwidth]{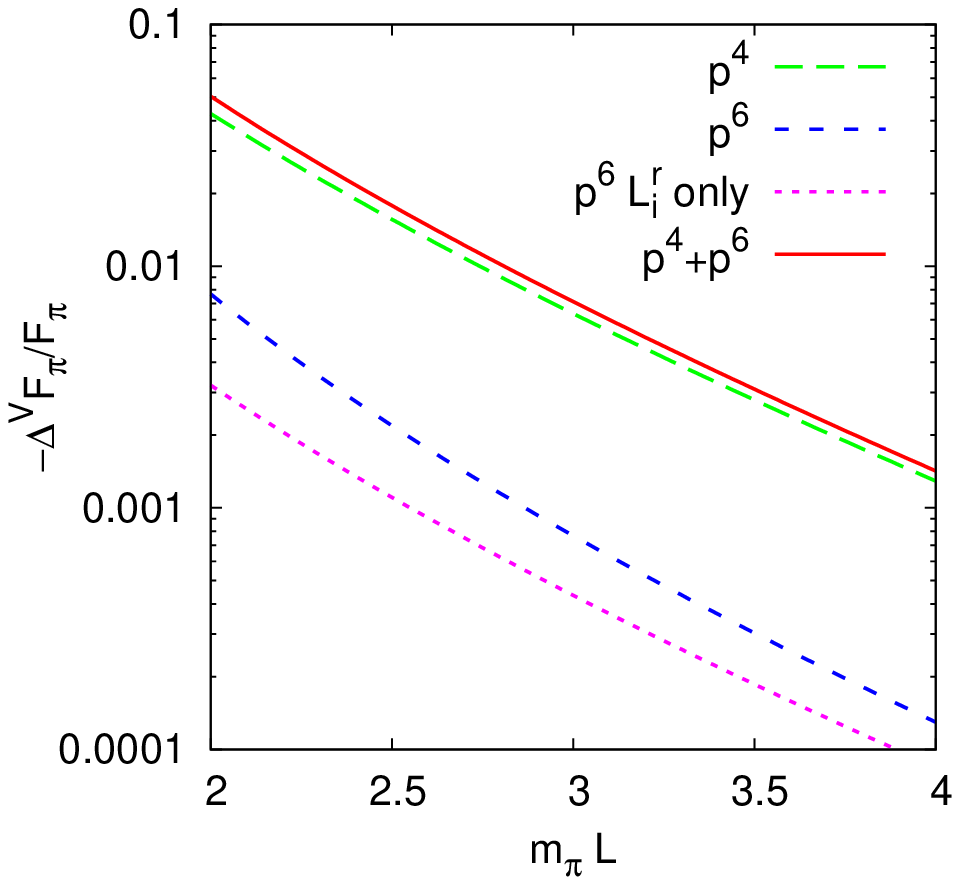}
\vskip-5mm
\centerline{(b)}
\end{minipage}
\caption{The finite volume corrections to the pion decay constant as a function
of $L$.
The other inputs can be found  in the text. 
Plotted is $-(F_\pi^{V}-F_\pi)/F_\pi$.
 (a) Comparison of the two- and three-flavour results. 
(b) The corrections for the three-flavour case showing
the $L_i^r$ dependent part separately.}
\label{figfpi}
\end{figure}

The decay constants can be calculated in a similar fashion. We reproduced
the known infinite volume two-loop results as well
as the one-loop results. We have a small disagreement with the partial
two-loop results in \cite{Colangelo:2005gd}. Numerical results for the pion
decay constants are shown in Fig.~\ref{figfpi}.

\begin{figure}[tb]
\begin{minipage}{0.49\textwidth}
\includegraphics[width=0.99\textwidth]{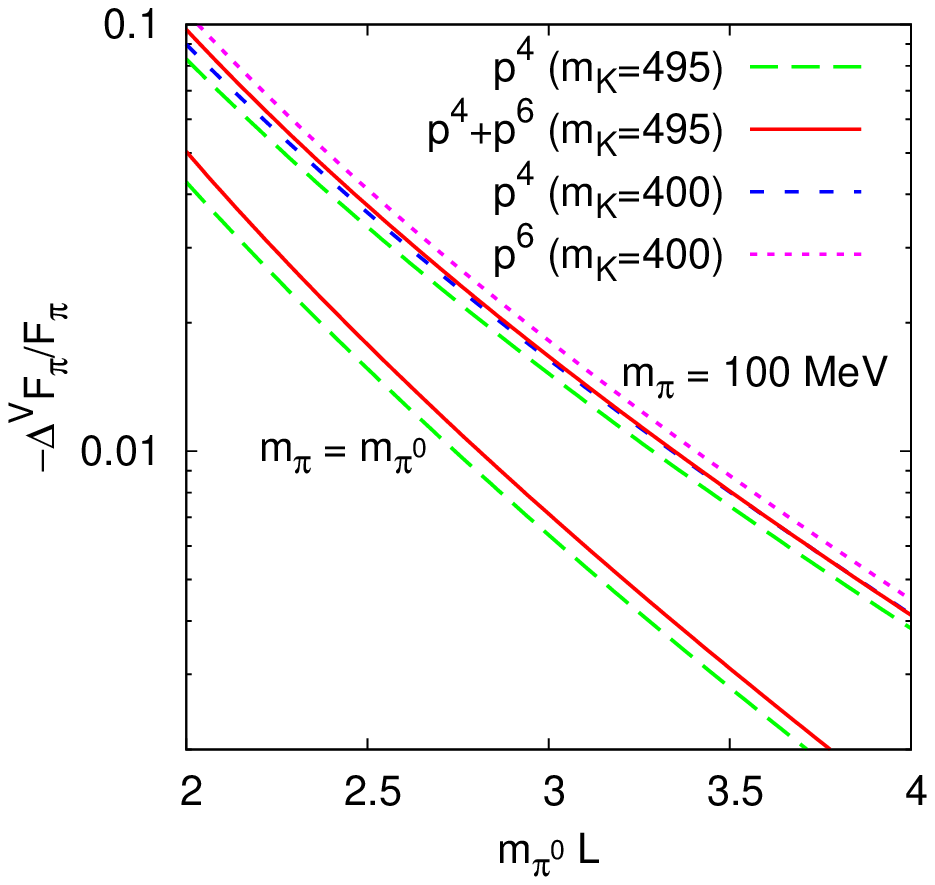}
\vskip-5mm
\centerline{(a)}
\end{minipage}
\begin{minipage}{0.49\textwidth}
\includegraphics[width=0.99\textwidth]{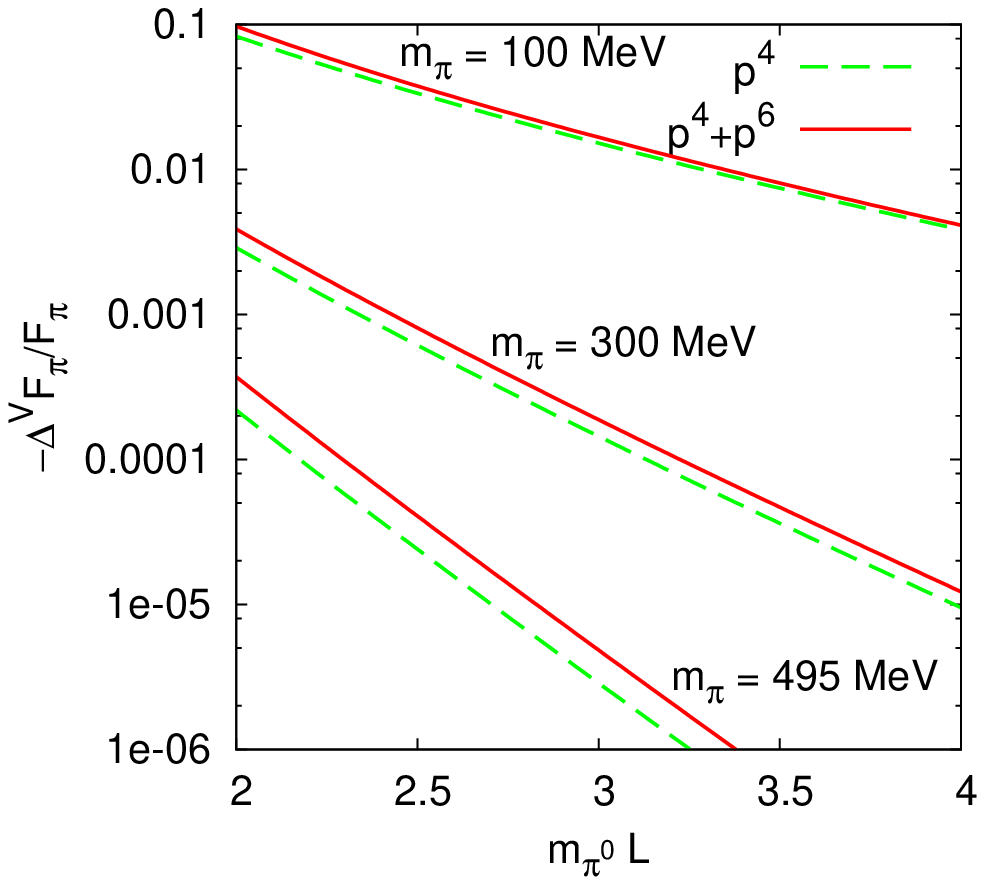}
\vskip-5mm
\centerline{(b)}
\end{minipage}
\caption{The finite volume corrections to the pion decay constant
for a number of mass cases.
Plotted is the quantity $-(F_\pi^{V}-F_\pi)/F_\pi$.
(a) Physical case and $(m_\pi,m_K) = (100,495)$ and $(100,400)$~MeV.
(b) $m_K=495$~MeV and $m_\pi=100,300,495$~MeV.
The size $L$ is given in units of the physical $\pi^0$ mass.}
\label{figfpi2}
\end{figure}

Some examples using the same LECs as above but varying input kaon
and pion masses with a calculated consistent value for $F_\pi$ and $m_\eta$
are shown in Fig.~\ref{figfpi2}. The method of calculating consistent
values is explained in \cite{Bijnens:2014dea}.

\section{Finite volume results for partially quenched three flavour ChPT}
\label{sec:FVPQChPT}

\subsection{Partial quenching}

\def\){\right)} 
\def\({\left(} 
\def\tr{\mbox{tr}}
\def\str{\mbox{str}}
\def\sdet{\mbox{sdet}}
\def\Dslash{{\rm D}\!\!\!\!/\,}

In the partially quenched case, we give different masses to the valence and
sea quarks. Valence quarks have flavour lines which connect to the external
fields. Sea quarks, on the other hand, appear only in closed loops. 
The distinction is crucial for comparison to partially quenched lattice
calculations. For the sea quarks, a (functional) fermion determinant appears
in the generating functional
\begin{equation}
\label{equationone}
Z=\int D[\psi\bar{\psi}A] 
\exp \( -S_{G}-\int \left[\bar{\psi}\(\Dslash+m\)\psi \right] \)
=\int
D[A]\exp\(-S_G\)\det\(\Dslash+m\)
\,
\end{equation}
when integrating over the anti-commuting fermionic Grassmann fields. Hence,
the exponential containing the gauge field action $S_G$ is modified by a
factor containing the fermion mass $m$.\footnote{The equations also hold for
several fermions, arranged in a column vector $\psi$, and a mass matrix $m$.} 
The (Monte-Carlo) evaluation of the determinant is computationally very
expensive, it is much cheaper to vary the valence quark mass only. Then, the
produced gauge field configurations do not need to be changed in the lattice
computation, but only the mass scale that is used to calculate the operator
expectation value.

An important issue for a partially quenched ChPT (PQChPT) calculation is the
question how to prevent quark fields in the valence sector from contributing
according to equation (\ref{equationone}) to the determinant. One way is given
by a construction, due to Morel \cite{Morel:1987xk}, that involves bosonic
spin-$1/2$ fields. By fixing the mass matrix $\tilde{m}$ of these ``ghost''
fields to the one used in the valence sector, the unwanted contribution
cancels exactly from the determinant as can be seen from
\begin{eqnarray}
Z&=&\int D[\psi\bar{\psi}\tilde{\psi}\bar{\tilde{\psi}}A] 
\exp \( -S_{G}-\int \left[\bar{\psi}\(\Dslash+m\)\psi+
\bar{\tilde{\psi}}\(\Dslash+\tilde{m}\)\tilde{\psi} \right] \)
       \nonumber      \\    &=&\int
D[A]\exp\(-S_G\)\frac{\det\(\Dslash+m\)}{\det\(\Dslash+\tilde{m}\)}
\,
\end{eqnarray}
with $\tilde{\psi}$ denoting the bosonic Dirac field.
This method, also called the supersymmetric formulation of PQChPT, has the
advantage that the partial quenching is completely performed by the
construction of the Goldstone manifold and the Lagrangian. Given that, the
Feynman-diagrammatic calculation then simply follows the ordinary rules of
Field Theory. It should be noted though on the side that this does not change
the fact that the partially quenched theory is no longer a proper Quantum
Field Theory in the strict sense: The supersymmetric formulation allows
rewriting partially quenched theory in terms of a local Lagrangian for
a statistical mechanical model - in violation of the spin-statistics theorem.

\newcommand{\Tr}{\mathrm{Tr}}
\newcommand{\Str}{\mathrm{Str}}
\newcommand{\diag}{\mathrm{diag}}
\newcommand{\nn}{\nonumber\\}
\newcommand{\lgl}{\langle}
\newcommand{\rgl}{\rangle}

The chiral group in the supersymmetric framework is formally extended to the graded
\begin{equation}
G = SU(n_\mathrm{val}+n_\mathrm{sea} | n_\mathrm{val})_L
\times SU(n_\mathrm{val}+n_\mathrm{sea} | n_\mathrm{val})_R
\end{equation}
for the case of $n_\mathrm{val}$ valence and $n_\mathrm{sea}$ quarks.
$G$ is then spontaneously broken to the diagonal subgroup
$SU(n_\mathrm{val}+n_\mathrm{sea} | n_\mathrm{val})_V$.
We have done our calculations in the flavour basis rather than in the meson basis.
We thus use fields $\phi_{ab}$ corresponding to the flavour content
of $q_a\bar q_b$. The mixing of the neutral eigenstates and the
integrating out of the singlet degree of freedom is taken
care of by using a more complicated propagator.
It is possible to use the same
method also in standard ChPT.

The corresponding Goldstone manifold is then parametrized by fields with generic structure
\begin{equation}
\label{SUSY_FieldMatrix}
\Phi =
\left(\begin{array}{ccc}
\Big[\;\;q_V\bar q_V\;\;\Big] & 
\Big[\;\;q_V\bar q_S\;\;\Big] &
\Big[\;\;q_V\bar q_B\;\;\Big] \\ \\ 
\Big[\;\;q_S\bar q_V\;\;\Big] &
\Big[\;\;q_S\bar q_S\;\;\Big] & 
\Big[\;\;q_S\bar q_B\;\;\Big]\\ \\
\Big[\;\;q_B\bar q_V\;\;\Big] & 
\Big[\;\;q_B\bar q_S\;\;\Big] &
\Big[\;\;q_B\bar q_B\;\;\Big]
\end{array}\right)\,
\label{pqmatr}
\end{equation}
where $V$ denotes valence, $S$ denotes sea and $B$ denotes the bosonic ghost quarks.
Note that the meson fields containing one single ghost quark only will
themselves obey fermionic, i. e. anticommuting, statistics.

The structure of the Lagrangian is similar to standard ChPT for a generic
number of flavours.
The lowest order Lagrangian is
\begin{equation}
\label{L2}
{\cal L}_2 = \frac{F_0^2}{4}\lgl u_\mu u^\mu + \chi_+\rgl\,. 
\end{equation}
At one-loop, the terms relevant to our work are given by
\begin{eqnarray}
\label{L4}
{\cal L}_4 &=&
 {\hat L}_0\,\lgl u^\mu u^\nu u_\mu u_\nu \rgl 
+{\hat L}_1\,\lgl  u^\mu u_\mu \rgl^2 
+{\hat L}_2\,\lgl u^\mu u^\nu \rgl \lgl u_\mu u_\nu \rgl
+{\hat L}_3\,\lgl (u^\mu u_\mu)^2 \rgl
\nn &&
+ {\hat L}_4\,\lgl u^\mu u_\mu \rgl \lgl \chi_+\rgl 
+ {\hat L}_5\,\lgl u^\mu u_\mu \chi_+ \rgl 
 +{\hat L}_6\,\lgl \chi_+ \rgl^2 
+ {\hat L}_7\,\lgl \chi_- \rgl^2
+ \frac{{\hat L}_8}{2}\,\lgl \chi_+^2 + \chi_-^2 \rgl 
+ \ldots\,.
\end{eqnarray}
The generalized Goldstone manifold is parametrized as
\begin{equation}
u\equiv\exp\left(i\Phi/(\sqrt{2}\hat F)\right)
\end{equation}
similar to the exponential representation in standard ChPT. For three physical flavours, it is a
$9\times9$ matrix with fermionic parts.
We have
furthermore introduced
\begin{eqnarray}
u_\mu &=& i\left\{
u^\dagger(\partial_\mu-i r_\mu)\,u -
u\,(\partial_\mu-i l_\mu)\,u^\dagger\right\},
\nonumber \\
\chi_\pm &=& u^\dagger\chi\,u^\dagger\pm u\,\chi^\dagger\,u\,.
\label{uquant}
\end{eqnarray}
The matrix $\chi$ is for this work restricted to
\begin{equation}
\chi = 2 B_0\,\mathrm{diag}(m_1,\ldots,m_9)\,
\end{equation}
with $m_i$ the quark mass of quark $i$ and $B_0$ a LEC. 
We have here $m_1=m_7, m_2=m_8, m_3=m_9$ as the valence
masses and $m_4,m_5,m_6$ as the sea quark masses.
Ordinary traces have been replaced by supertraces, denoted by
$\lgl~\rgl$, defined in terms of the ordinary ones by
\begin{equation}
\Str \left(\begin{array}{cc} A & B \\ C & D \end{array}\right)
=\Tr\,A - \Tr\,D\,.
\end{equation}
$B$ and $C$ denote the fermionic blocks in the matrix.
The supersinglet $\Phi_0$, generalizing the $\eta'$, is integrated out
to account for the axial anomaly as in standard ChPT,
implying the additional condition
\begin{equation}
\label{tracezero}
\lgl \Phi \rgl = \Str\,(\Phi) = 0\,.
\end{equation}
However, as mentioned above, we will work in the flavour basis enforcing
the constraint (\ref{tracezero}) via the propagator.

A calculation in PQChPT has to be performed using a larger set of operators
since no further reduction by means of Cayley-Hamilton relations can be
performed. The three-flavour PQChPT Lagrangian (equation (\ref{L4})) thus
has 11 LECs for PQChPT.

%The LECs for standard three flavour ChPT
%are related to those of three flavour PQChPT via
%\begin{equation}
%L_1^r = \hat L_1^{r} + \hat L_0^{r}/2,\qquad
%L_2^r = \hat L_2^{r} + \hat L_0^{r},\qquad
%L_3^r = \hat L_3^{r} - 2\,\hat L_0^{r},
%\end{equation}
%and $L_i^r=\hat L_i^r$ for the others.
%%
%Note that a numerical value for $\hat L_0$ cannot be obtained by experiment,
%but can be determined only via PQQCD lattice simulations or modelling.

An additional comment is that the divergences for PQChPT are directly
related to those for $n_{sea}$-flavour ChPT \cite{Bijnens:1999hw}
when all traces are replaced by supertraces. This can be argued using
the formal equivalence of the equations of motion used or via the
replica trick \cite{Damgaard:2000gh}.

This method was used in the pioneering work for actual calculations
in partially quenched ChPT \cite{Bernard:1993sv}. More details can be found
in \cite{Sharpe:2000bc,Sharpe:2001fh}. In those references you can also find
a detailed derivation of the double poles that appear in the propagators
for neutral particles.

\subsection{Quark flow}

An alternative method to do (partially) quenched calculations in ChPT
is to use the quark flow method introduced in \cite{Sharpe:1992ft}.
It basically consists of working in the flavour basis and keeping all
flavour lines. The removal of the singlet degree of freedom is now done
via extra terms in the neutral propagator. In the end one checks
which flavour lines are connected to external fields or operators and those
are given the corresponding valence quark flavour.
The remaining ones are to be summed over the sea quark flavours. 
This was generalized to the two-loop diagrams in our calculation.

\subsection{Results}

The analytical calculations were performed using both methods, supersymmetric
and quark flow, with results in full agreement.
We also reproduced the known infinite volume results
\cite{Bijnens:2004hk,Bijnens:2005ae,Bijnens:2006jv} and the one-loop finite
volume expressions \cite{Aubin:2003mg,Aubin:2003uc}.
The expressions for the different mass cases reduce to each other and
to the unquenched case when taking the relevant mass limits.

For the LECs we use the results of the $L^r_i$ of the recent fit
\cite{Bijnens:2014lea} and we set the additional LEC $L_0^r=0$.
The scale is set to $\mu=0.77$~GeV
and the lattice size $L$ we choose as
that $M L=2$ for $M=0.13$~GeV. For the lowest order pion decay constant we
use $F_0=87.7$~MeV. The lowest order kaon mass we fix to 450~MeV.

\begin{figure}[tb]
\begin{minipage}{0.49\textwidth}
\includegraphics[width=0.99\textwidth]{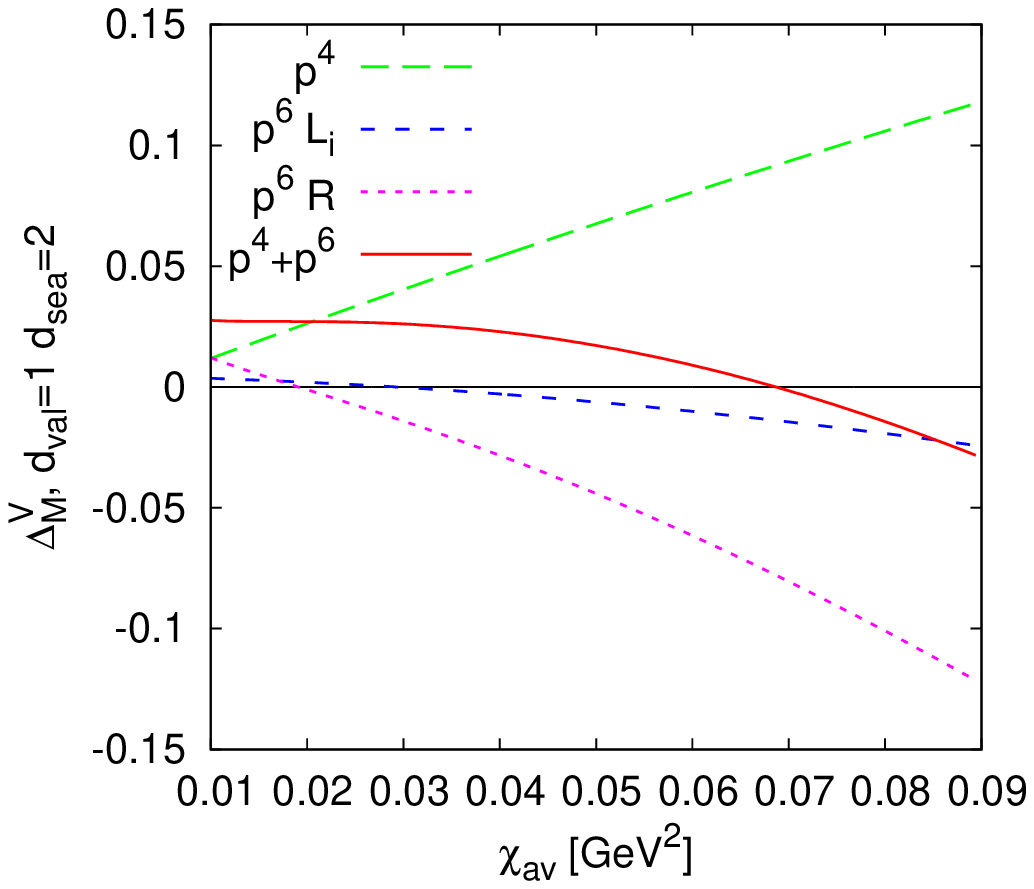}
\vskip-5mm
\centerline{(a)}
\end{minipage}
\begin{minipage}{0.49\textwidth}
\includegraphics[width=0.99\textwidth]{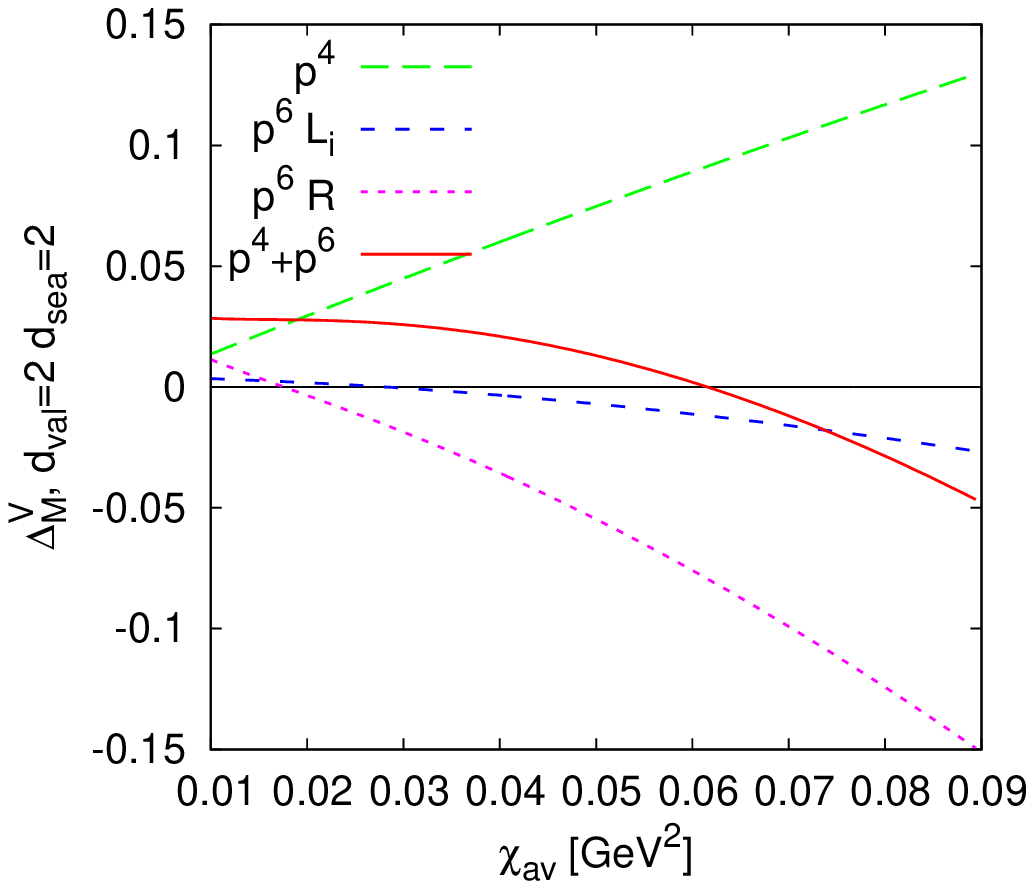}
\vskip-5mm
\centerline{(b)}
\end{minipage}
\begin{minipage}{0.49\textwidth}
\includegraphics[width=0.99\textwidth]{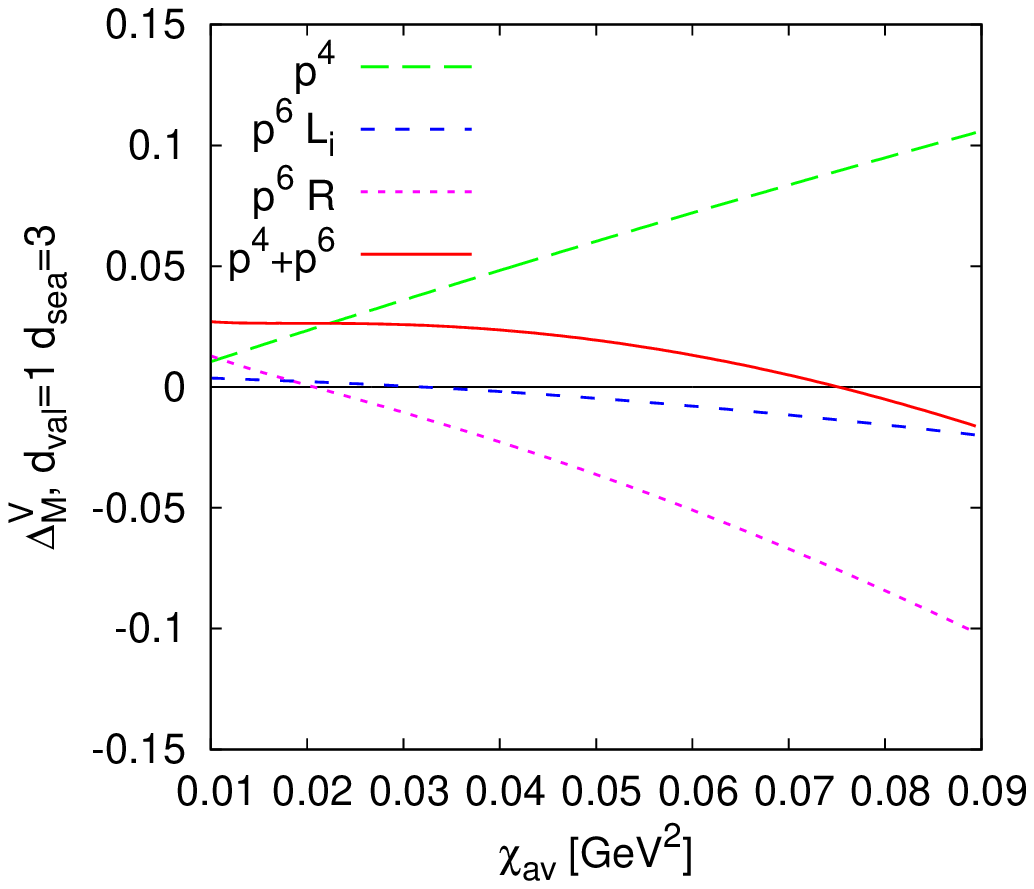}
\vskip-5mm
\centerline{(c)}
\end{minipage}
\begin{minipage}{0.49\textwidth}
\includegraphics[width=0.99\textwidth]{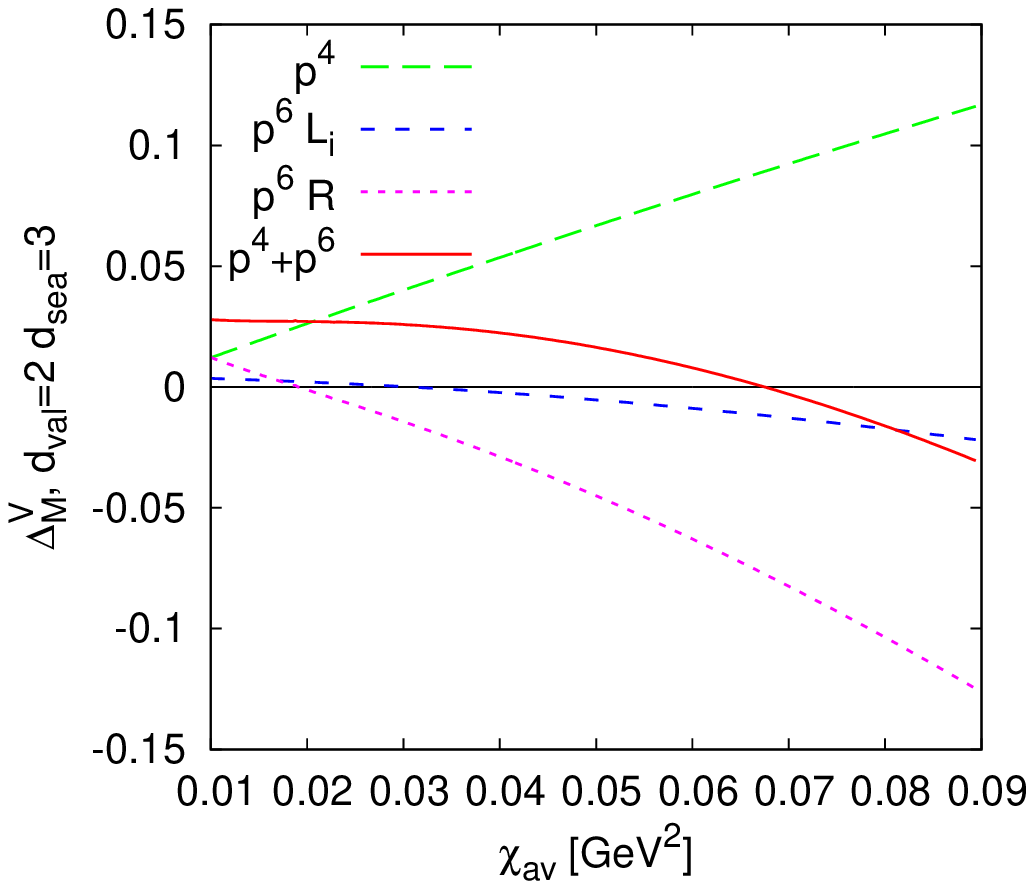}
\vskip-5mm
\centerline{(d)}
\end{minipage}
\caption{\label{figmpiPQ} The corrections for the pion mass relative
to the lowest order mass as a function
of the average up and down sea quark mass via $\chi_\mathrm{av}$.
When isospin breaking is included the ration of up to down quark mass
is chosen to be 1/2.
(a) The isospin limit in sea and valence,
(b) Isospin breaking in the valence sector only.
(c) Isospin breaking in the sea sector only.
(d) Isospin breaking in both sectors.}
\end{figure}

\begin{figure}[tb]
\begin{center}
\includegraphics[width=0.49\textwidth]{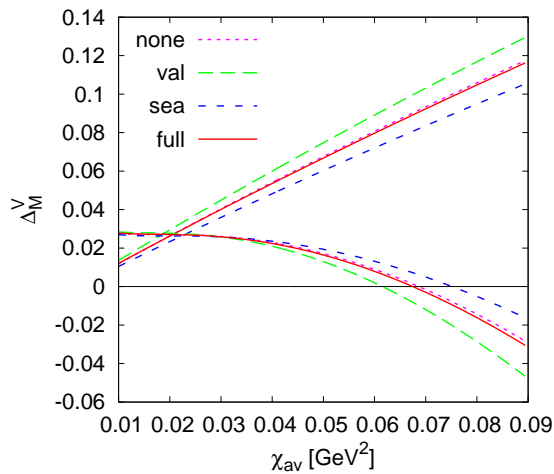}
\end{center}
\caption{\label{figiso} Comparing the finite volume correction for
the meson masses for the cases with no isospin breaking (none),
only in the valence sector (val), only in the sea sector (sea) and in both
(full) for the meson mass squared. The upper curves are the $p^4$, the bottom the $p^4+p^6$ results.}
\end{figure}

As an example we plot the pion mass for a number of cases
where we keep the lowest order (valence) pion mass at 130~MeV
but vary the sea quark masses to get a sea pion mass varying from
100 to 300~MeV. The sea strange quark is fixed at a mass slight above the
valence sea mass. The variation with the sea pion mass squared
$\chi_\mathrm{av}$ is shown in Fig.~\ref{figmpiPQ}.
We show the cases for up-down mass equal for both sea and valence,
different for valence only, different for sea only and both different.
The cancellation between the up-down mass differences between sea and valence
effects is accidental. It does not happen for the decay constant.
The four different cases are compared in Fig.~\ref{figiso}.

\begin{figure}[tb]
\begin{minipage}{0.49\textwidth}
\includegraphics[width=0.99\textwidth]{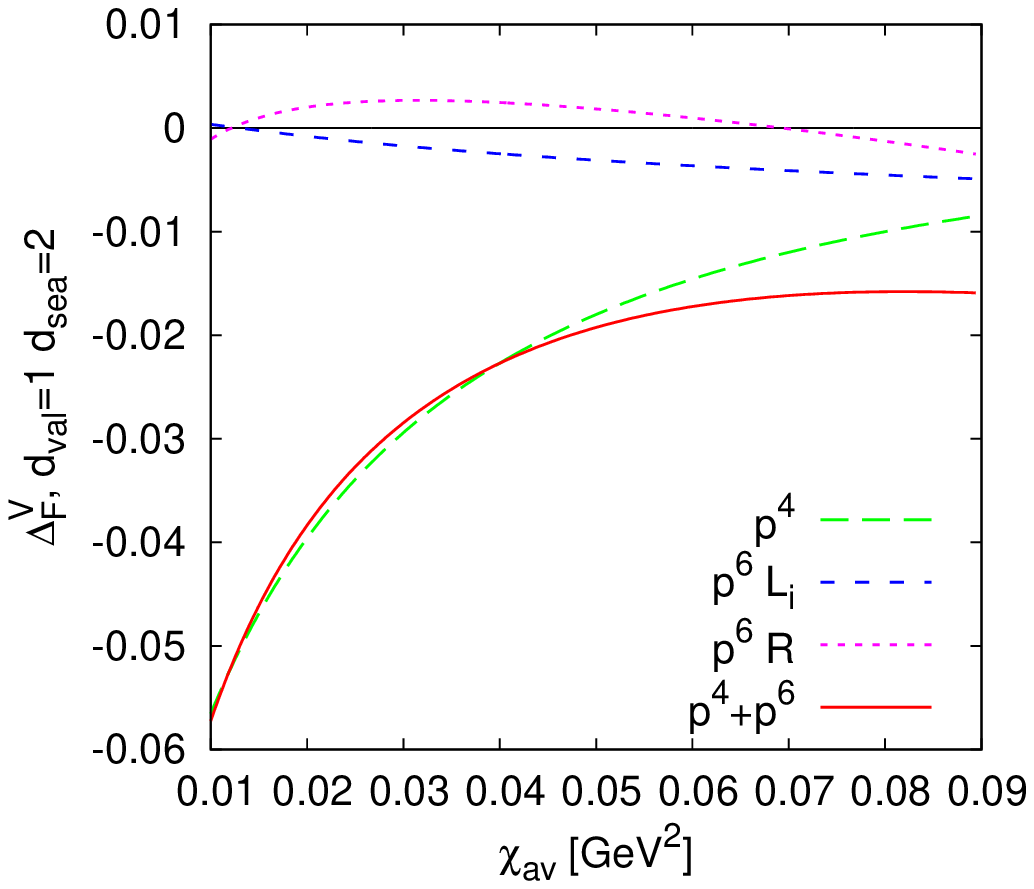}
\vskip-5mm
\centerline{(a)}
\end{minipage}
\begin{minipage}{0.49\textwidth}
\includegraphics[width=0.99\textwidth]{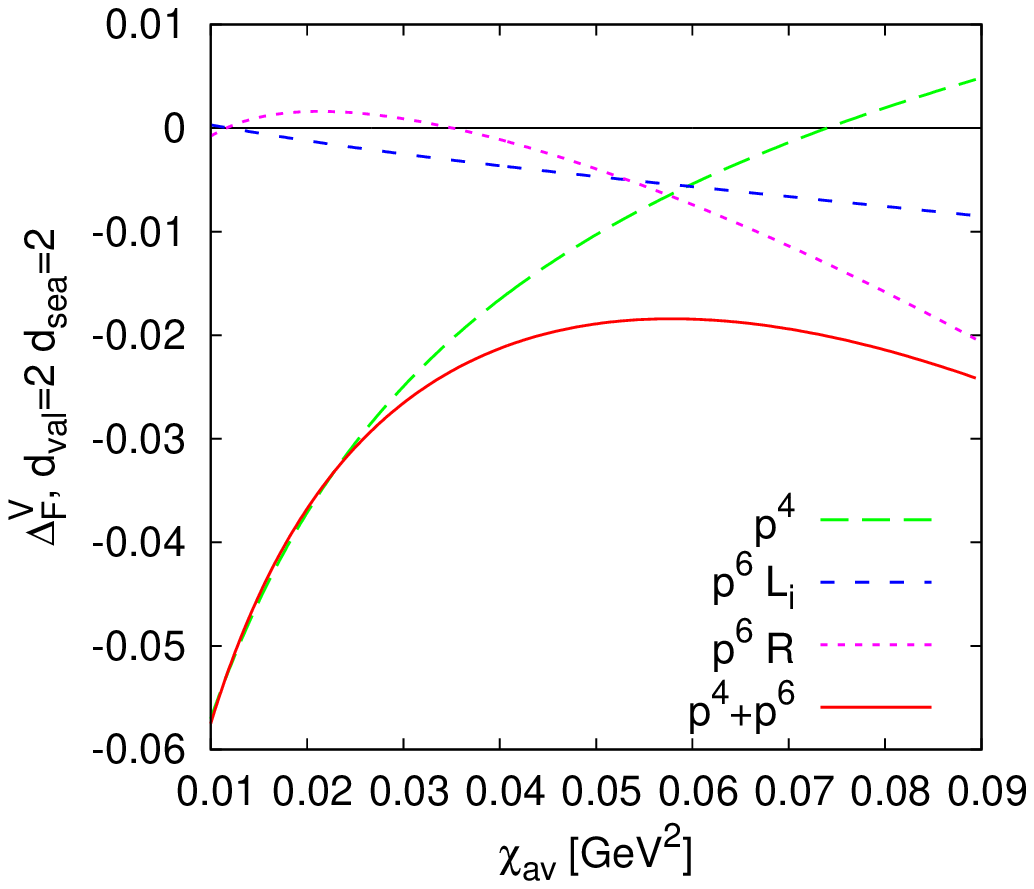}
\vskip-5mm
\centerline{(b)}
\end{minipage}
\begin{minipage}{0.49\textwidth}
\includegraphics[width=0.99\textwidth]{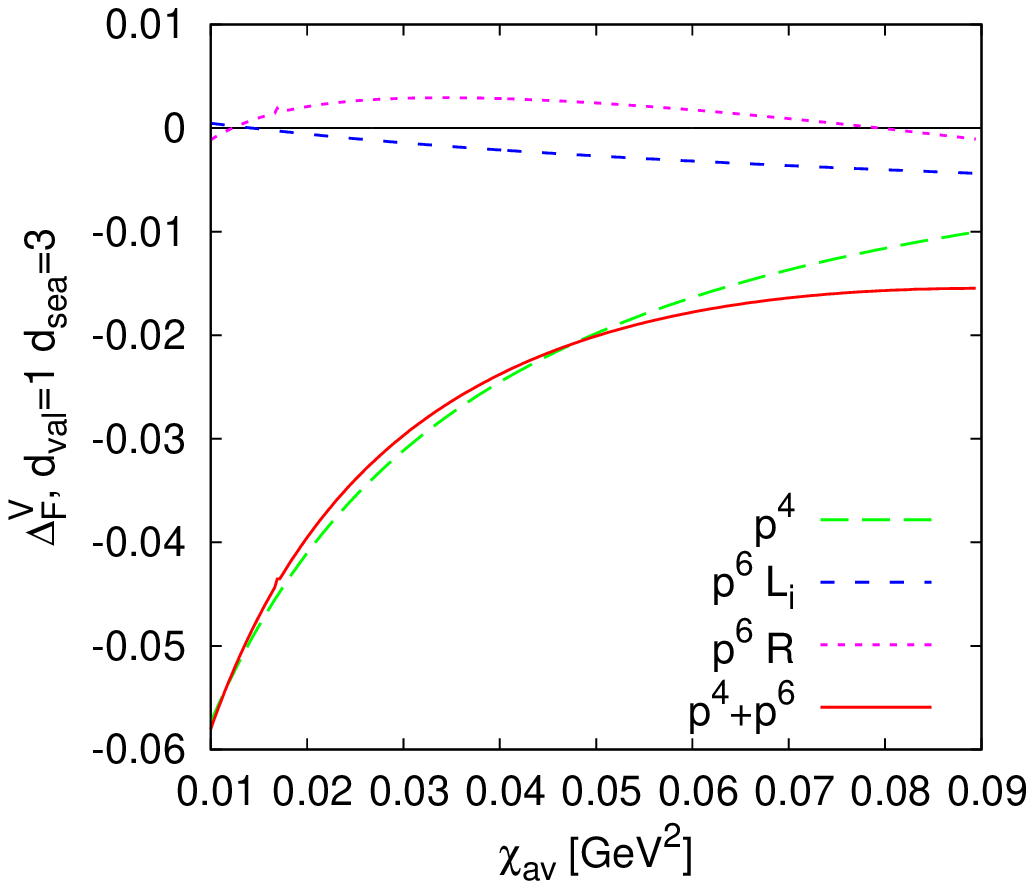}
\vskip-5mm
\centerline{(c)}
\end{minipage}
\begin{minipage}{0.49\textwidth}
\includegraphics[width=0.99\textwidth]{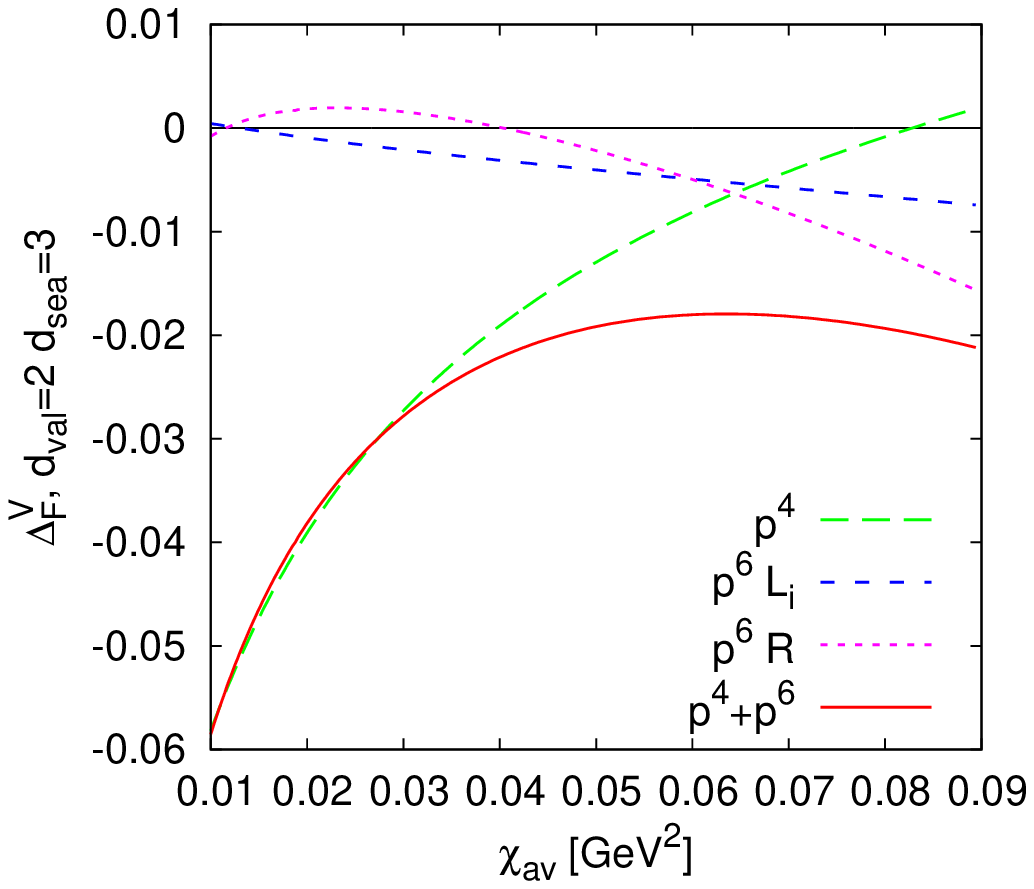}
\vskip-5mm
\centerline{(d)}
\end{minipage}
\caption{\label{figfpiPQ} The corrections for the pion decay constant relative
to its lowest order value as a function
of the average up and down sea quark mass via $\chi_\mathrm{av}$.
(a) The isospin limit.
(b) Isospin breaking in the valence sector.
(c) Isospin breaking in the sea sector.
(d) Isospin breaking in both sectors.}
\end{figure}

The same cases for the pion decay constant are shown in Fig.~\ref{figfpiPQ}.
The cancellation which happened for the masses is not present here.
Figures for a number of other cases can be found in the paper.

\section{Update: QCD-like theories}

The finite volume corrections for masses, decay constants and the
quark-antiquark vacuum expectation value in the effective field
theory for QCD-like theories have been recently calculated as well.
The extension to the partially quenched case was done in the same
work \cite{Bijnens:2015xba}. This work can be seen as an extension
of our work in ChPT discussed above and of the
unquenched infinite volume results of \cite{Bijnens:2009qm}.

\acknowledgments
We thank the organizers for a very pleasant and well-run conference.
This work is supported in part by the Swedish Research Council grants
621-2011-5080 and 621-2013-4287.

\end{document}